\documentclass[aps,prl,twocolumn,amssymb,amsfonts,floatfix]{revtex4-1}
\usepackage{graphicx}
\usepackage[section]{placeins}
\usepackage{amsmath}
\usepackage{bm}
\usepackage{longtable}
\usepackage{times}
\usepackage{amssymb}
\usepackage{color}

\newcommand{\beq}{\begin{equation}}
\newcommand{\eeq}{\end{equation}}
\newcommand{\bea}{\begin{eqnarray}}
\newcommand{\eea}{\end{eqnarray}}

\begin{document}
\title{Trispectrum reconstruction of non-Gaussian noise}
\author{Guy Ramon}
\email{gramon@scu.edu}
\affiliation{Department of Physics, Santa Clara University, Santa Clara, CA 95053}
\begin{abstract}

Using a qubit to probe non-Gaussian noise environments is theoretically studied in the context of classical random telegraph processes. Protocols for control pulses are developed to effectively scan higher noise correlations, offering valuable information on the charge environment of the qubit. Specifically, the noise power spectrum and trispectrum are reconstructed simultaneously for a wide range of qubit-fluctuator coupling strengths, demonstrating the method's robustness. These protocols are readily testable in various qubit systems with well-developed quantum control, including quantum dot spins, superconducting qubits and NV centers in diamond.
\end{abstract}


\maketitle


\emph{Introduction.} The decoherence experienced by all solid-state qubits is largely determined by
their immediate environment that typically includes fluctuating charges and spins of nuclei or local electron. While unavoidable interactions with the environment continue to limit the number of qubits that can be coherently manipulated and entangled, recent studies attempt to harness qubit susceptibility to local fluctuations by transforming them into high-resolution sensitive probes of their environment. Valuable on its own \cite{Degen_RMP17}, noise spectroscopy is anticipated to help mitigating decoherence more effectively, by, e.g., modifying the physical hosting system or adjusting the qubit control schemes.

In recent years a growing body of works across a wide range of quantum systems has been devoted to exploring strategies to characterize environmental noise by measuring the dynamics of properly initialized and controlled qubits \cite{Szan}.
Operating a qubit under a sequence of dynamical decoupling (DD) pulses was shown to establish a simple relation between the measured qubit signals and the noise power spectrum \cite{Alvarez_PRL11,Bylander_Nature11,Almog_JPB11,Yuge_PRL11}. Dynamically decoupling a qubit inflicted by Gaussian phase noise effectively generates a frequency-domain filter determined by the pulse sequence \cite{Cywinski_PRB08}. For periodic sequences with a fixed pulse interval $\tau$, the filter is sharply peaked at a frequency $f=1/2\tau$ and its odd harmonics, allowing one to scan the noise spectrum by subjecting the qubit to sequences with varying pulse intervals. This method of DD-based noise spectroscopy (DDNS) has been used to reconstruct environmental noise in various qubit platforms, including trapped ions \cite{Biercuk_Nature09,Kotler_Nature11}, superconducting circuits \cite{Bylander_Nature11,Sung_ArXiv19}, semiconductor quantum dots (QDs) \cite{Medford_PRL12,Chan_PRAp18,Kawakami_PNAS16}, phosphorous donors in silicon \cite{Muhonen_Nature14}, and NV centers in diamond \cite{Bar-Gill_Nature12,Staudacher_Science13,Romach_PRL15}.

DDNS has been limited by several approximations in the underlying theory and method implementation, namely, the qubit is assumed to undergo pure dephasing by coupling longitudinally to the noise, and the noise is taken to be both classical and Gaussian. Several recent theoretical studies have put forth qubit control protocols that allow, in principle, to extend the applicability of DDNS methods to non-Gaussian \cite{Norris_PRL16}, and quantum \cite{Paz-Silva_NJP16,Paz-Silva_PRA17} noise environments.

Focusing on the assumption of noise Gaussianity, we note that a Gaussian process is most often the result of a collection of many uncorrelated (or weakly correlated) processes, e.g., when the qubit interacts with a large environment, so that all higher correlation functions factorize into products of one- and two-point correlators. This picture breaks down when high-resolution sensing of small environments strongly coupled to the qubit is pursued, or when environmental dynamics with strong and nontrivial spatial correlations is present due to interactions between its constituents \cite{Clerk_RMP10,Lisenfeld_Nature15}. Furthermore, solid-state devices are commonly afflicted with low-energy excitations, such as slowly switching two-level fluctuators (TLFs) that are responsible for $1/f$ noise \cite{Paladino_PRL02,Paladino_RMP14}, which is inherently non-Gaussian, thus in many realistic scenarios a complete noise characterization must include higher correlations and their respective noise cumulants.

In this letter we study reconstruction of polyspectra --- the Fourier transforms of multi-point correlators --- of a classical Random Telegraph Noise (RTN) process, building on an approach proposed by Norris {\it et al.} \cite{Norris_PRL16}. Using carefully chosen sequences of DD pulses, we adapt the method to reconstruct the RTN power spectrum and trispectrum. The central role of charge noise in limiting qubit coherence has been long established in various platforms such as Josephson qubits \cite{Astafiev_PRL04,Martinis_PRL05}, and QDs \cite{Jung_APL04, Dial_PRL13}, and was more recently reaffirmed in both \cite{Beaudoin_PRB15} Si/SiGe \cite{Kawakami_PNAS16,Yoneda_Nature18,Connors_ArXiv19}, and GaAs \cite{Malinowski_PRL17,Cerfontaine_ArXiv19} QD spin qubits.
Moreover, RTN sources are the quintessential testbed for non-Gaussian noise spectroscopy protocols, owing in part to the ability to control the non-Gaussianity probed by the qubit by tuning the coupling strength to switching rate ratio, $\eta$. As $\eta$ increases, pronounced non-Gaussian behavior is formed, exhibiting, e.g., plateaus in the qubit decay signal \cite{Bergli_NJP09,Ramon_PRB15}. We provide explicit control protocols that are readily accessible in various solid-state qubit systems, explain the numerical challenges associated with polyspectra reconstruction, and offer their resolution.

Before discussing our results, we mention that other approaches to quantify non-Gaussian noise have been considered, including approximate resummation of the cumulant series \cite{Cywinski_PRA14}, appropriate for the special case of quadratic coupling to a Gaussian noise (e.g., at an optimal working point), and a non-perturbative spectroscopy method that was demonstrated for non-Gaussian discrete magnetic noise, strongly coupled to the qubit \cite{Kotler_PRL13}.


\emph{Formalism.} At pure dephasing, the qubit-TLF Hamiltonian reads ${\cal H}=b_z(t) \hat{\sigma}_z/2$, where $b_z (t)=v \xi (t)$, $v$ is the coupling strength, $\xi(t)=\pm 1$ represents the RTN stochastic variable switching between the two states with an average rate $\gamma$, and we assume for simplicity that the average times spent at each state are equal (i.e., symmetric TLF). Qubit dephasing at readout  time $T$ is manifested in the off-diagonal elements of its density matrix, ${\hat \rho}_Q(T)$, which is constructed from the evolution operator including any control pulses. Qubit coherence is then quantified by:
\beq
W(T) \equiv \frac{{\left\langle  +  \right|{{\hat \rho }_Q}(T)\left|  -  \right\rangle }}{{\left\langle  +  \right|{{\hat \rho }_Q}(0)\left|  -  \right\rangle }} = \left\langle {{e^{ - i\phi (T)}}} \right\rangle  \equiv {e^{ - \chi (T)}},
\eeq
where $\langle \cdot \rangle$ denotes averaging over realizations of the classical RTN (or partial trace of the environmental degrees of freedom for quantum noise). For any noise, a formal solution to the attenuation factor, $\chi (T)$, can be written in terms of a cumulant expansion \cite{Kubo,Cywinski_PRB08}
\beq
\chi (T) =  - \sum\limits_{k = 1}^\infty  {\frac{{{{( - iv)}^k}}}{{k!}}{C_k}(T)},
\label{chi}
\eeq
where the $k$th cumulant of the noise, $C_k(T)$, includes all connected diagrams of that order and can be written in terms of the noise moments. We assume qubit evolution under a sequence of $n_p$ ideal $\pi$ pulses about the $x$ or $y$ axes defined by the switching function: $f_T (t) =\sum_{j=0}^{n_p} (-1)^k \Theta (t-t_j) \Theta (t_{j+1}-t)$,
where we take $t_0=0$ and $t_{n_p +1}=T$. For a given switching function, the $k$th cumulant is found as
\beq
{C_k}(T) \! = \! \int_0^T \!\!\! {{f_T}({t_1})d{t_1}}  \cdots \!\int_0^T \!\!\! {{f_T}({t_k})d{t_k}} {A_k}\left( {{t_1}, \ldots ,{t_k}} \right),
\label{Ckt}
\eeq
where $A_k\left( {t_1}, \ldots ,{t_k} \right)$ includes all the $j$-point correlation functions, $\langle \xi (t_1) \cdots \xi (t_j) \rangle$,  up to $j=k$. For stationary noise, the correlators depend only on time separations, $\tau_{j} \equiv t_{j+1}-t_1$,  $j\in 1, \ldots , k-1$, and the cumulants can be evaluated in Fourier space:
\beq
{C_k}(T) \! = \!\! \int_{\mathbb{R}^{k -\! 1}} \!\! \frac{d\vec{\omega}_{k \! -1}}{{\left( {2\pi} \right)}^{k \!-\! 1}} \! \prod_{j=1}^{k-1} \!\tilde f_T ({\omega_j}) \tilde f_T \!\left(\!-\Sigma \vec{\omega}_{k - 1} \right) \! S_{k - 1}\! \left(\vec{\omega }_{k -\! 1} \!\right)\!,
\label{Ck}
\eeq
where ${\tilde f_T}(\omega)$ is the Fourier transform of the switching function, known as the filter function, ${\vec \omega _k} \equiv \left( \omega _1, \ldots ,\omega_k \right)$, $\Sigma \vec{\omega}_k \equiv \omega _1 + \ldots + \omega _k$, and we introduced the polyspectra \cite{Norris_PRL16}:
\beq
S_k\left( {\vec \omega }_k \right) =  \int_{\mathbb{R}^k} \!{d{{\vec \tau }_k}{e^{ - i\left( {{{\vec \omega }_k } \cdot {{\vec \tau }_k}} \right)}}{A_{k+1}}\!\left( {{{\vec \tau }_k}} \right)}.
\label{Sk}
\eeq
For Gaussian noise all higher noise moments factorize to products of lower moments and their respective cumulants vanish. Furthermore, if the Gaussian noise has zero mean, $\langle \xi(t) \rangle=0$, it is fully characterized by its two-point correlation function, or equivalently by its Fourier transform--- the power spectral density (PSD), also referred to as the first spectrum \cite{Kogan_96}. In contrast, a complete account of a non-Gaussian noise must include all cumulants and their respective polyspectra.

We now apply this formalism to treat the case of a single RTN. The noise generated by a symmetric TLF has zero mean and all odd noise moments and cumulants vanish, thus the resulting attenuation factor given by Eq.~\ref{chi} includes only decay with no phase contribution. Starting from the two- and four-point correlators: $\langle \xi (t_1) \xi (t_2) \rangle=e^{-2\gamma |t_1 -t_2|}$, and $ \langle \xi(t_1) \xi (t_2) \xi (t_3) \xi (t_4) \rangle=e^{-2 \gamma (t_4-t_3+t_2-t_1)}$, $t_1 \leq t_2 \leq t_3 \leq t_4$, with corresponding permutations for other time orderings \cite{Galperin_04,Cywinski_PRB08}, we have $A_2(t_1,t_2) = \langle \xi(t_1) \xi(t_2) \rangle$, and
\begin{eqnarray*}
&& A_4 (t_1,t_2,t_3,t_4) = \langle \xi(t_1) \xi(t_2) \xi(t_3) \xi(t_4) \rangle - A_2(t_1,t_2) \times \\ && A_2(t_3,t_4)- A_2(t_1,t_3) A(t_2,t_4) -A_2(t_1,t_4) A_2(t_2,t_3).
\end{eqnarray*}
The resulting PSD and trispectrum are found respectively as
\beq
S_1 (\omega) = \frac{4\gamma}{4\gamma^2 +\omega^2},
\label{S1}
\eeq
and \cite{coupling} (see section I of the supplemental material \cite{Supp})
\beq
S_3 \left( \vec{\omega}_3 \right) \!=\!  - 16\gamma \frac{ 48\gamma ^4 \!+ \! 4\gamma ^2 \sum_{i \leq j} \omega _i \omega _j + \omega _1 \omega _2 \omega _3 \Sigma \vec{\omega}_3}{\prod\limits_{i=1}^3 \left( 4\gamma ^2 + \omega _i^2 \right)\left(4\gamma ^2 + (\Sigma \vec{\omega}_3)^2 \right)}.
\label{S3}
\eeq
For free induction decay (FID) or simple pulse sequences, such as periodic dynamical decoupling (PDD) and Carr-Purcell-Meiboom-Gill (CPMG), the second and fourth cumulants of a single classical RTN were calculated analytically in time domain using Eq.~(\ref{Ckt}) \cite{Cywinski_PRB08,Ramon_PRB15}, matching the results obtained from Eq.~(\ref{Ck}) by direct integration of Eqs.~(\ref{S1}) and (\ref{S3}). Polyspectra of any classical noise are highly symmetric and are fully specified by their values within a frequency space known as the principal domain \cite{Norris_PRL16,Chandran_IEEE94}. Whereas the PSD possesses a single (even) symmetry that defines its principal domain as all non-negative frequencies, the trispectrum is invariant under 48 operations, as detailed in section II of the supplemental material.

DDNS protocols rely on the application of pulse sequences whose filter functions are characterized by a fundamental frequency, $2\pi/T$, and its harmonics, effectively acting as frequency combs \cite{Alvarez_PRL11}. Recently scrutinized \cite{Szankowski_PRA18}, the accuracy of this so-called delta approximation improves significantly by repetition of the base sequences, effectively extending the measurement time to $MT$, where $M$ is the number of repetitions \cite{Ajoy_PRA11}. Extending these protocols to non-Gaussian noise spectroscopy relies on the ability of pulse sequences to form multidimensional frequency combs necessary for polyspectra reconstruction \cite{Norris_PRL16}. Approximating the filter functions with multidimensional dirac combs allows us to replace the frequency integrals in Eq.~(\ref{Ck}) with summations, truncated at frequencies where either the filter function or the polyspectrum (or both) are sufficiently small. The resulting approximate second and fourth cumulants are:
\beq
C_2 (MT) \approx \frac{M}{T} \! \sum_{n \in {\cal D}_1} \! \Omega_1(n) \left|  \tilde{f}_T \! \left( \frac{2\pi n}{T}\right)\right|^2 \! S_1 \!\left(  \frac{2\pi n}{T} \right)
\label{C2}
\eeq
\bea
C_4 (MT) & \approx & \frac{M}{T^3} \!\! \sum_{\vec{n}_3 \in {\cal D}_3} \!\! \Omega_3\!\left(\vec{n}_3\right) \prod_{j=1}^3 \tilde{f}_T \left( \frac{2\pi \vec{n}_3(j)}{T}\right) \times \nonumber \\
&& \tilde{f}_T \left( -\frac{2\pi \Sigma \vec{n}_3}{T}\right) S_3 \!\left(  \frac{2\pi \vec{n}_3}{T} \right),
\label{C4}
\eea
where ${\cal D}_1= \{ 0,1, \ldots ,n^{\rm max}\}$ denotes the discretized principal domain of the PSD, truncated at frequency $2\pi n^{\rm max}/T$, and the corresponding multiplicity is $\Omega_1 (n)=2-\delta_{n,0}$. Likewise, $\vec{n}_{3}$ denotes all three-tuple integers up to $n_3^{\rm max}$ within the trispectrum discretized and truncated principal domain, ${\cal D}_3$, with corresponding  multiplicity, $\Omega_3$ \cite{Supp}.

The final step needed in order to form a finite set of equations connecting the polyspectra to measurable qubit signals, is to truncate the cumulant series in Eq.~(\ref{chi}), thereby treating non-Gaussian terms perturbatively. In the context of RTN spectroscopy this approximation seems to suggest that we are limited to weak coupling, $\eta=v/\gamma \ll 1$. Nevertheless, the noise reconstruction procedure described below can be used to extract valuable information on the charge environment even when it is strongly coupled to the qubit.


\emph{DDNS Protocols.} The original DDNS protocol proposed by \'{A}lvarez and Suter \cite{Alvarez_PRL11} utilized a set of 2-pulse CPMG sequences with variable sequence times: $T_i=T,T/2, \ldots T/n^{\rm max}$, scanning the spectrum with a resolution of $\omega_{\rm min}=2\pi/T$ up to spectral bound of $\omega_{\rm max}=2\pi n^{\rm max}/T$. Non-Gaussian DDNS requires a large number of distinct sequences to support the reconstruction of multidimensional grids of frequency points. Following Norris {\it et al.}, we construct a pool of base sequences comprising Mixed-order Concatenated DD (MCDD) segments with orders between 0 to 5. FID segments are included to avoid full refocusing of static noise, which is important for polyspectra reconstruction as the number of points with zero frequency becomes substantial. All base sequences have a fixed time $T=256 \delta=16/\gamma$, where $\delta$ is the time resolution providing a theoretical upper bound of $\pi/\delta$ for the frequency cutoff, and time is scaled with inverse RTN switching rate, $\gamma$ \cite{Tgamma}. Other requirements imposed on the MCDD sequences included in the pool are detailed in the section III of the supplemental material. Setting harmonics bounds at $n^{\rm max}=32$, and $n^{\rm max}_3=8$, the reconstructed PSD and trispectrum are truncated at frequencies $\pi/4\delta$, and $\pi/16\delta$, respectively. Retaining only the second and fourth cumulants, the set of linear equations that is formed by inserting Eqs.~(\ref{C2}) and (\ref{C4}) into Eq.~(\ref{chi}) can be cast as
\beq
\vec{\chi}=A \left( \begin{array}{c} \vec{S}_1 \\ \vec{S}_3 \end{array} \right).
\label{chiN}
\eeq
Our aforementioned harmonics bounds correspond to $n_{\rm PSD}=33$ and $n_{\rm TRI}=285$ frequency points in the PSD and trispectrum, respectively, thus $n_{\rm seq}=318$ sequences and corresponding signal measurements are needed. The reconstruction matrix, $A$, connects these measured signals with the PSD and trispectrum. We stress that restricting the reconstruction to frequencies within the principal domains is not only advantageous in reducing the number of equations to be solved --- it is, in fact, essential in order to avoid singularities in $A$.

\begin{figure}[!bt]
\vspace*{-0.65 cm}
\includegraphics[scale=0.36]{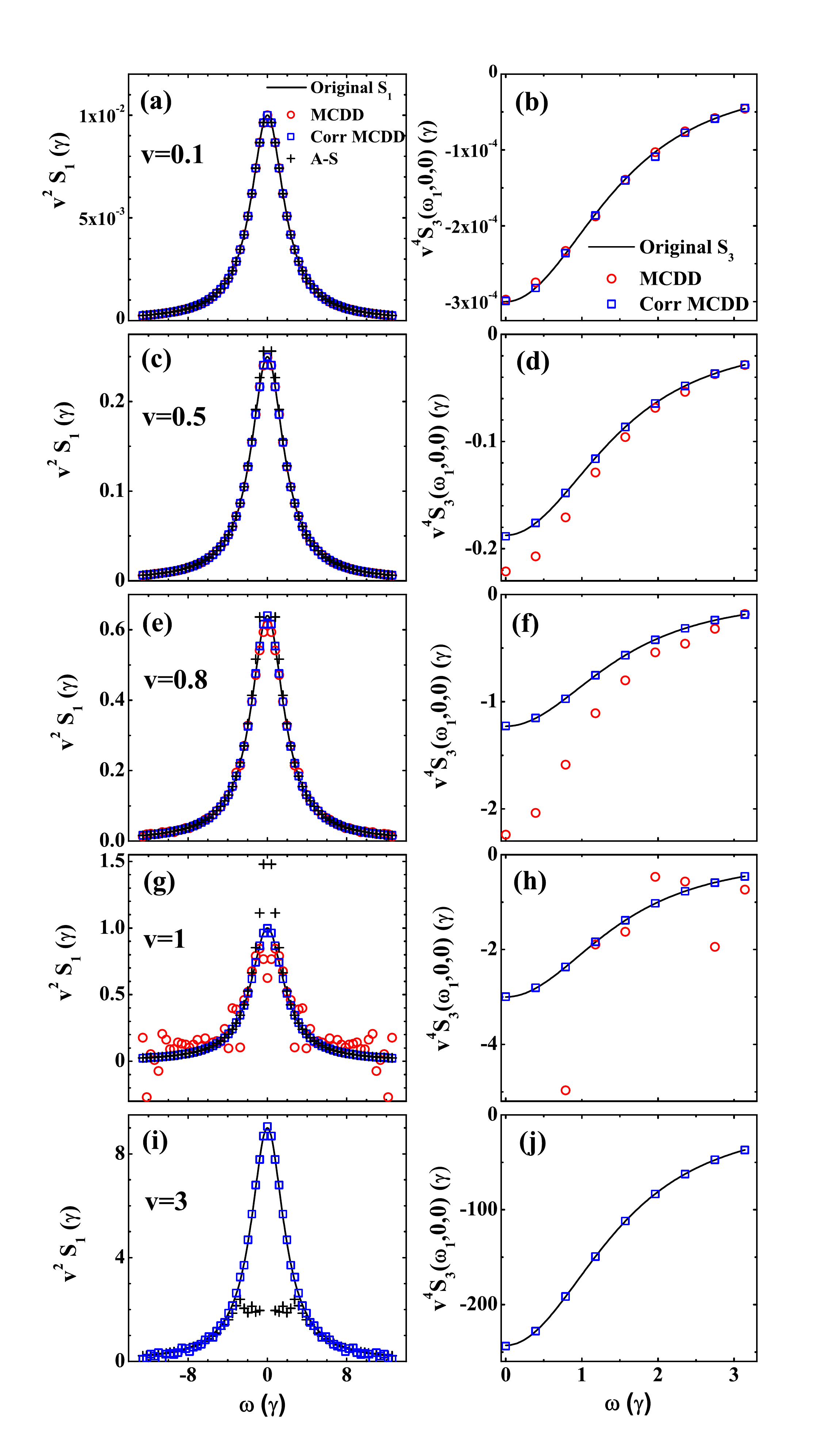}
\vspace*{-0.65 cm}
\caption{(Color online) RTN reconstruction of PSD (panels a, c, e, g, i) and trispectrum cut, $\omega_2=\omega_3=0$ (panels b, d, f, h, j). Starting at top row, coupling strengths are $v=0.1$, $0.5, 0.8, 1$, and 3. Solid lines correspond to the theoretical spectra, Eqs.~(\ref{S1}) and (\ref{S3}), red circles depict reconstructions with sets of MCDD sequences, optimized for each $v$, and blue squares show corrected results based on a-priori knowledge of he RTN parameters (see main text). At $v=3$ (panels (i) and (j)), uncorrected polyspectra reconstruction with only the first two cumulants is no longer practical. Black pluses in the PSD panels depict Gaussian reconstruction with the \'{A}lvarez-Suter protocol, which is inadequate at strong coupling (see panel (i)). Frequencies are measured in units of RTN switching rate, $\gamma$. }
\label{Fig1}
\end{figure}

\emph{Reconstruction Results.} Several factors limit our ability to reconstruct non-Gaussian noise by introducing errors into Eq.~(\ref{chiN}): (i) truncation of the cumulant expansion, (ii) inaccuracy of the delta approximation in the spectroscopic formulas, Eqs.~(\ref{C2}) and (\ref{C4}), (iii) frequency cutoffs in the truncated principal domains, and (iv) numerical errors in the matrix inversion due to large condition numbers. Whereas the first error predominantly depends on the noise non-Gaussianity, dictated by the RTN parameters, the other three errors can be substantially reduced by appropriate selection of control sequences. Crucially, however, sequences that minimize errors due to discretization and frequency truncation tend to provide similar spectral filtering, resulting in numerical instabilities. Moreover, at weak coupling, where contributions from the neglected higher cumulants are insignificant, trispectrum reconstruction is nevertheless challenging as it requires very stringent error thresholds on the second cumulant, since its related matrix elements are much larger than those associated with the fourth cumulant.

We have developed a layered algorithm for optimized selection of sets of control sequences that balances between these conflicting requirements, enabling us to obtain faithful reconstructions of the PSD and trispectrum of single RTNs over a wide parameter range. Our sequence selection is primarily based on maximizing the accuracy of the cumulants evaluation, given predefined finite sets of frequency points in ${\cal D}_1$ and ${\cal D}_3$. Minimizing both absolute (for each sequence) and relative (within the entire sequence set) errors, as well as ensuring low condition numbers for the reconstruction matrix, $A$, are all crucial for successful reconstruction, as detailed in section III of the supplemental material. The resulting set of control sequences are recorded in $A$, and their corresponding qubit attenuation factors, $\chi(T)$, are calculated exactly for a given RTN, using a transfer matrix method \cite{Ramon_PRB12,Ramon_PRB15}. Fig.~\ref{Fig1} demonstrates PSD and trispectrum reconstructions for RTN with several coupling strengths. Polyspectra of order $n$ are plotted with a prefactor of $v^{n+1}$, to represent their weighted contribution in the reconstruction matrix.

\begin{figure}[!bt]
\vspace*{0.1 cm}
\includegraphics[scale=0.3]{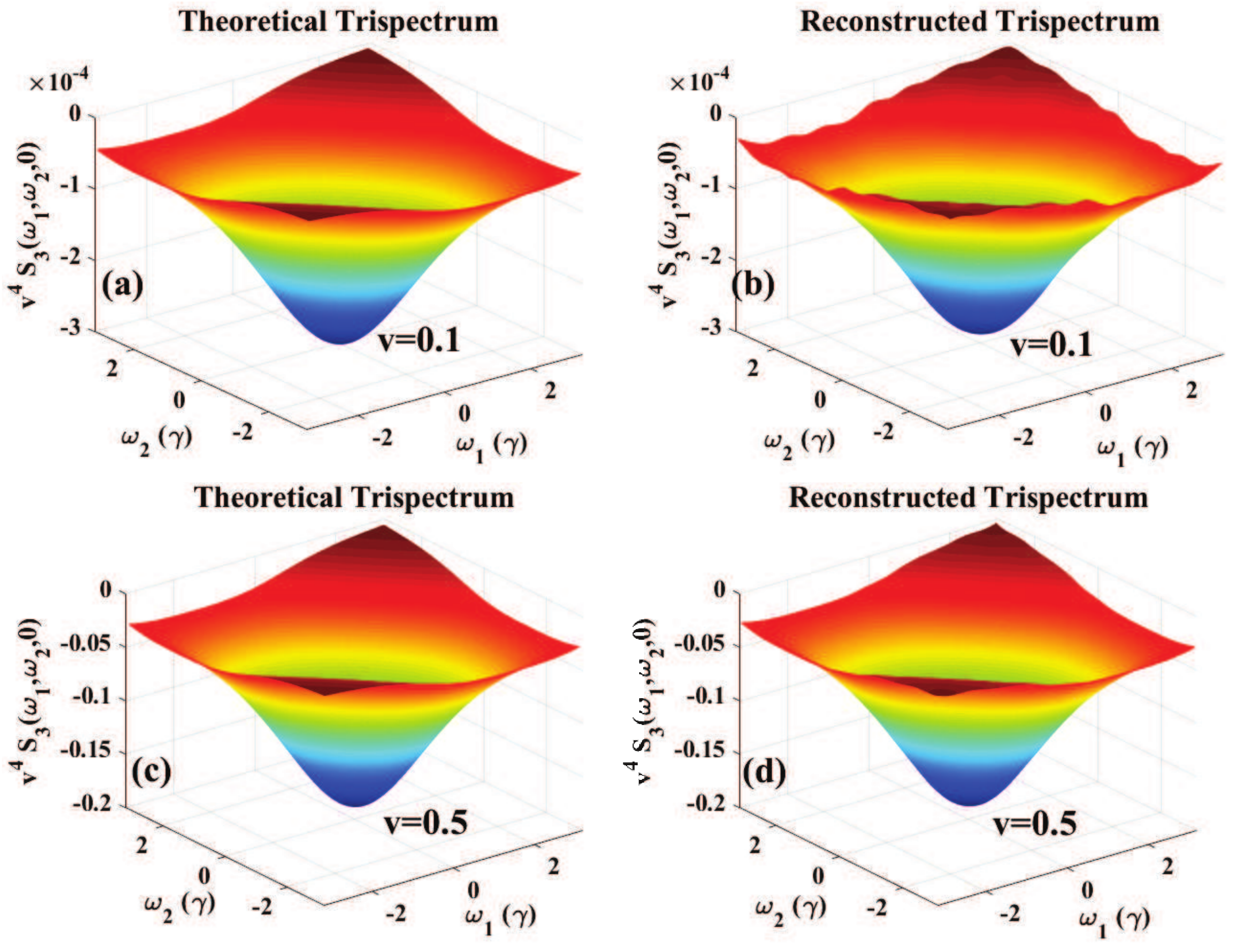}
\vspace*{-0.2 cm}
\caption{(Color online) Theoretical and reconstructed RTN trispectra cuts at $\omega_3=0$ for two coupling strengths. Spline interpolation was used to smoothen the reconstructed values in plots (b) and (d). The same sequence set, optimized for $v=0.1$ was used for both reconstructions.}
\label{Fig2}
\end{figure}
As coupling strength increases, non-Gaussianity becomes more pronounced and errors due to the cumulant series truncation dominate. We disentangle this (physical) error from the other (numerical) errors by applying a correction term to the attenuation factors used in Eq.~(\ref{chiN}) that accounts for the truncated cumulants. The resulting corrected reconstructions are depicted by blue squares in Fig.~\ref{Fig1} and their excellent agreement with the theoretical spectra shows that all other errors have been successfully minimized through our sequence selection procedure. We stress that while the cumulant correction term requires a-priori knowledge of the RTN parameters and is thus not directly applicable for experimentally measured qubit signals afflicted by an unknown noise source, our preliminary work suggests that with a reasonable overhead one can employ a feedback loop to simultaneously optimize noise reconstruction and identify the unknown RTN parameters. Since the set of control sequences optimized for weak coupling reconstruction satisfies the more stringent second cumulant error requirements, we expect and observe that the same protocol provides adequate reconstruction at stronger couplings, as long as cumulant error correction is applied for $v \gtrsim 0.5$. This is demonstrated in Fig.~\ref{Fig2}, where, unlike the results shown in Fig.~\ref{Fig1}, we use a single set optimized for $v=0.1$, for trispectra reconstruction at larger coupling strengths (see section III of the supplemental material for additional details).

\begin{figure}[!bt]
\vspace*{-0.5 cm}
\includegraphics[scale=0.22]{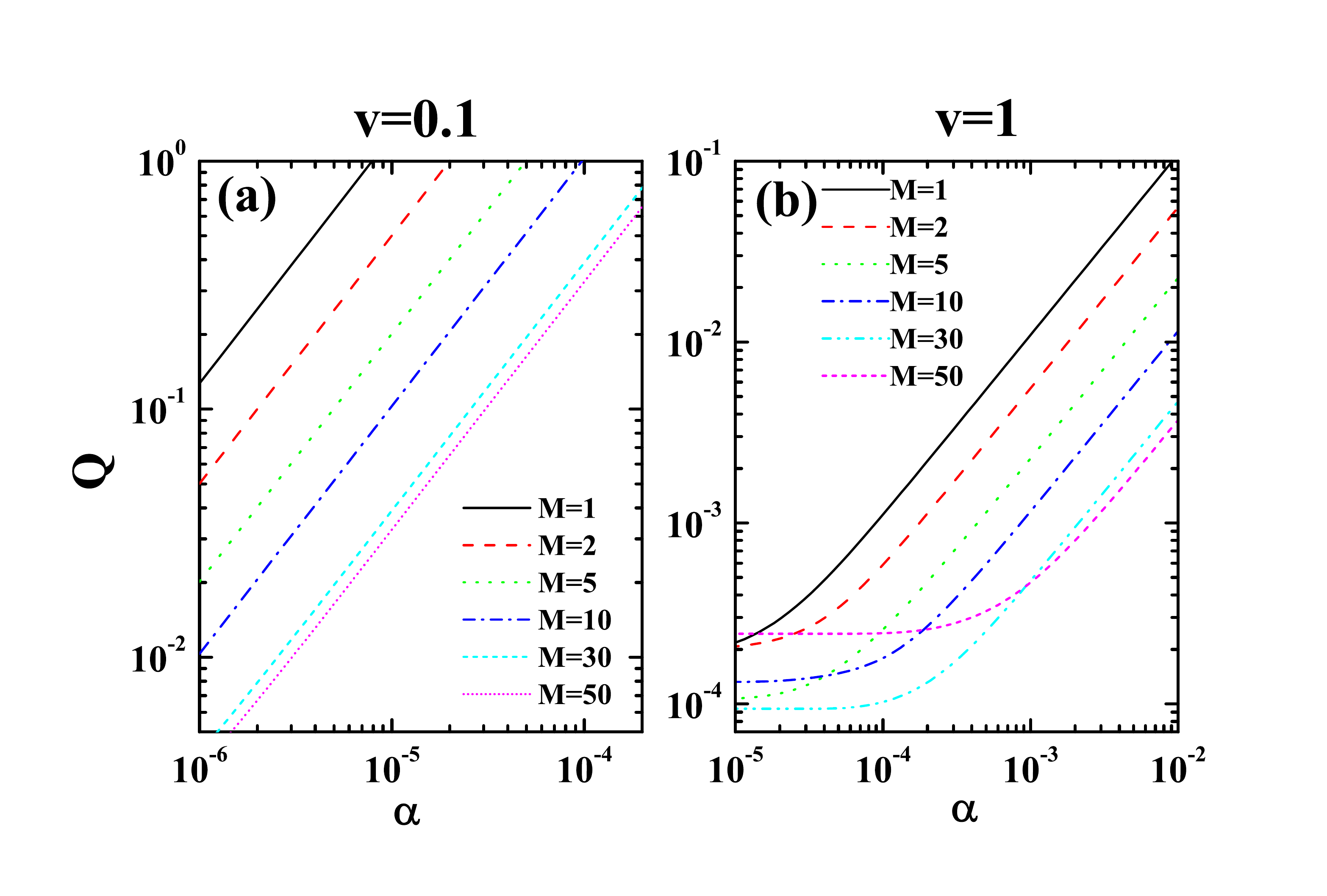}
\vspace*{-0.7 cm}
\caption{(Color online) Trispectrum reconstruction quality, ${\cal Q}$, vs.~relative measurement noise amplitude, $\alpha$, with several base sequence repetitions for RTN coupling strengths of: (a) $v=0.1$; (b) $v=1$. Sequence sets are optimized for each $M$ and $v$ value. In panel (b) reconstructions are corrected to account for truncated polyspectra higher than $S_3$.}
\label{Fig3}
\end{figure}
We studied the robustness of our noise reconstruction against measurement errors and the number of base sequence repetitions, $M$ \cite{repetition}. In Fig.~\ref{Fig3} we plot the reconstruction quality ${\cal Q}$, defined as
\beq
{\cal Q} \equiv \frac{1}{n_{\rm TRI}} \sum_{\vec{n}_3 \in {\cal D}_3} \left|\frac{  S_3 \left( \frac{2 \pi  \vec{n}_3}{T} \right)- S_3^{\rm rec} \left( \frac{2 \pi  \vec{n}_3}{T} \right)}{S_3 \left( \frac{2 \pi \vec{n}_3}{T} \right)}  \right|,
\label{Q}
\eeq
against the measurement noise amplitude, $\alpha$, relative to the qubit signal, for several $M$ values and two coupling strengths (see section IV of the supplemental material for more details on our measurement error simulations and protocol robustness). Here, we have employed sequence sets optimized for each given $M$ and $v$. At weak coupling, the dominant factor limiting reconstruction quality is cumulant errors due to the delta approximation. With increased $M$, this approximation becomes more accurate, resulting in consistent improvement in ${\cal Q}$. Interestingly, at larger coupling strengths, after applying the cumulant correction term, the reconstruction quality is limited by condition number instabilities, which tend to be stronger at larger $M$, where the filter function peaks are narrower. As shown in Fig.~\ref{Fig3}(b), increasing $M$ in this case is not always beneficial, even in the idealized scenario of perfect $\pi$-pulses shown here. Using a pulse sequence set optimized for given $v$ and $M$ to reconstruct polyspectra of RTN under different conditions always results in larger cumulant errors and consequently poorer reconstruction.

\emph{Conclusions} We developed control sequence protocols for trispectrum reconstruction of RTN sources. These protocols are shown to be robust over a wide range of RTN parameters, and can be used to identify and characterize single RTN sources in a variety of qubit platforms. We plan to extend this work in subsequent studies by developing control protocols that will map two and more RTN sources, and treat quantum TLFs in the strong coupling regime \cite{Abel_PRB08,Wold_PRB12}. Non-Gaussian noise spectroscopy can also be used to shed light on the role of TLF interactions in generating 1/f noise \cite{Lisenfeld_Nature15,Muller_PRB15,Ding_ArXiv18}. The current work can be viewed as complementary to a very recent experimental demonstration of non-Gaussian noise spectroscopy that utilized a flux superconducting qubit as a probe \cite{Sung_ArXiv19}. Whereas the non-Gaussian signatures there were observed in the phase evolution of the qubit's coherence, captured to leading order by the bispectrum, the non-Gaussianity of our RTN process is encoded to leading order in the trispectrum, as part of the signal decay. The multilayered sequence selection algorithm that we utilize to resolve numerical instabilities inherent in the original proposal \cite{Norris_PRL16} also differs from the maximum likelihood approach taken in \cite{Sung_ArXiv19}.

\emph{Acknowledgements} The author wishes to thank {\L}ukasz Cywi\'{n}ski for useful discussions. This work was supported by the National Science Foundation Grant no.~DMR 1829430.

\bibliographystyle{amsplain}

\begin{thebibliography}{99}
\bibitem{coupling} Our definition of the noise cumulants in Eq.~(\ref{chi}) excludes the coupling strength, $v$, and the resulting noise correlators and polyspectra should be multiplied by $v^k$ for correct evaluation of their weight in the cumulant expansion. Here, $k$ ($k-1$) corresponds to the $k$-point correllator ($k-1$ polyspectrum).
\bibitem{Tgamma} We expect that a simple preliminary measurement protocol can be carried out to calibrate the measurement time $T$ with the average switching rate of the RTN at hand.
\bibitem{repetition} Our analysis does not include pulse errors, but for most scenarios it is expected that pulse errors will become more dominant as the number of pulses increases. From this perspective, it is desirable to keep $M$ as low as possible, while maintaining a reasonably accurate delta-approximation.
\end{thebibliography}

\pagebreak
\onecolumngrid
\hspace*{-2cm}
\includegraphics{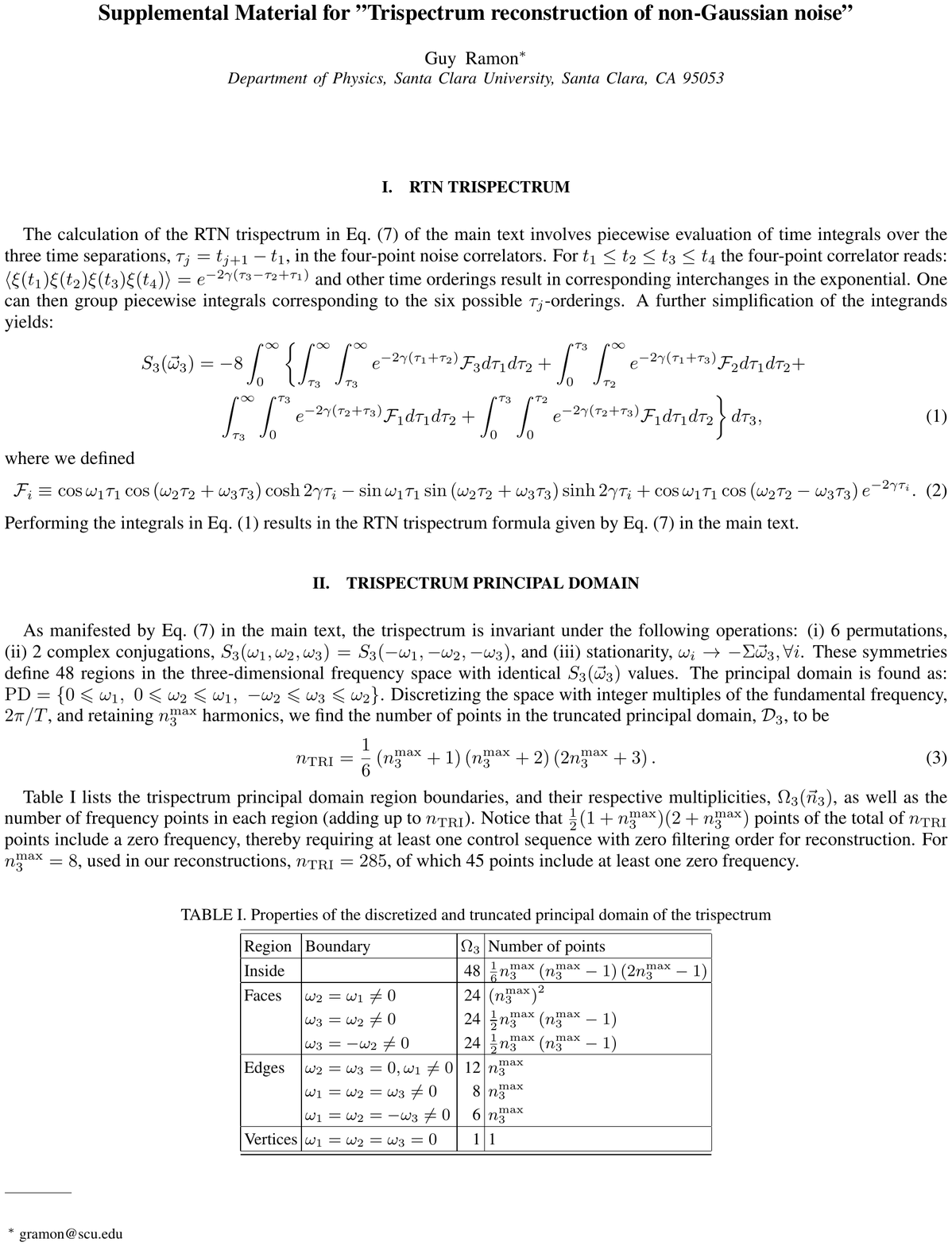}
\hspace*{-2cm}
\includegraphics{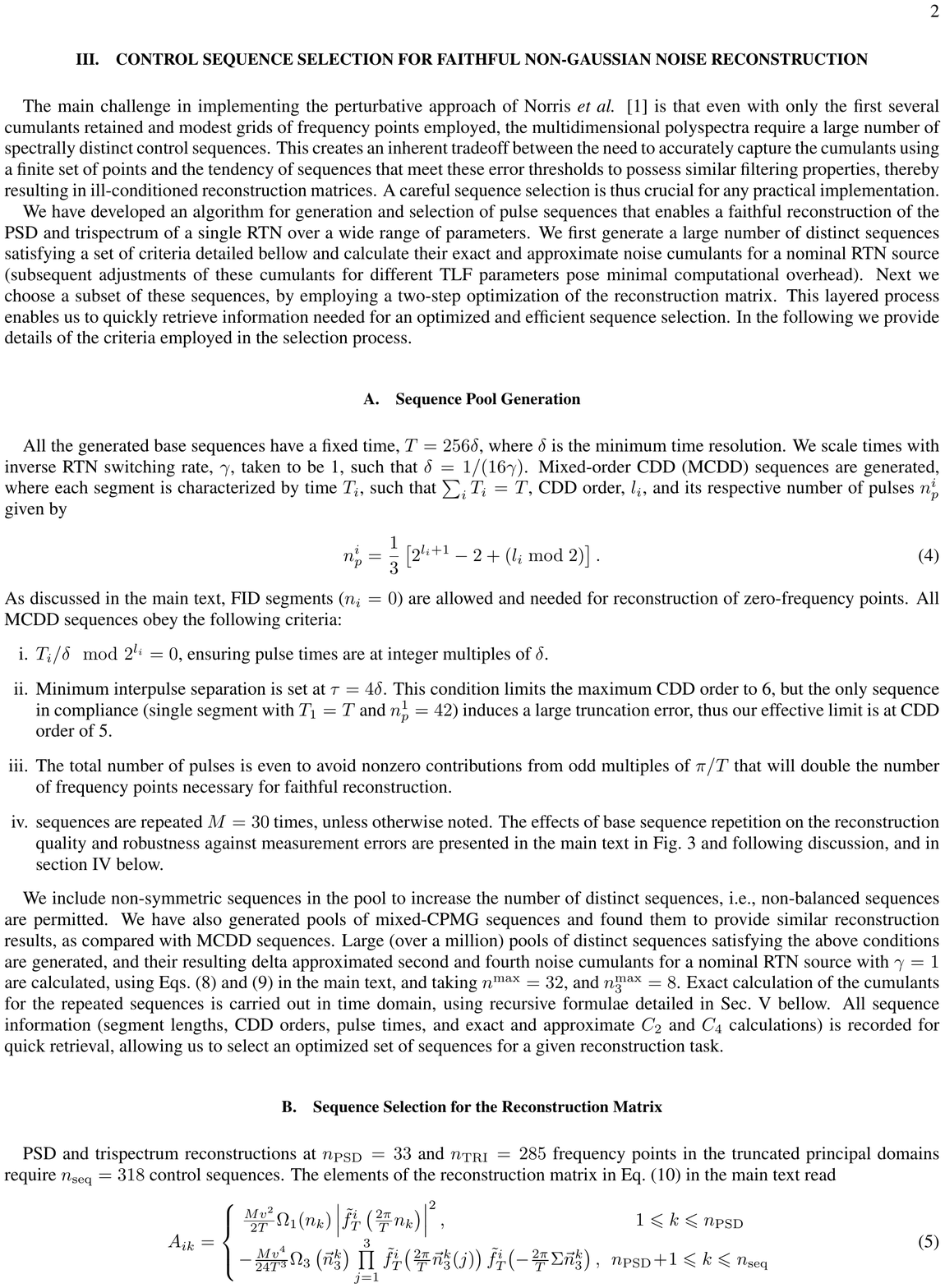}
\hspace*{-2cm}
\includegraphics{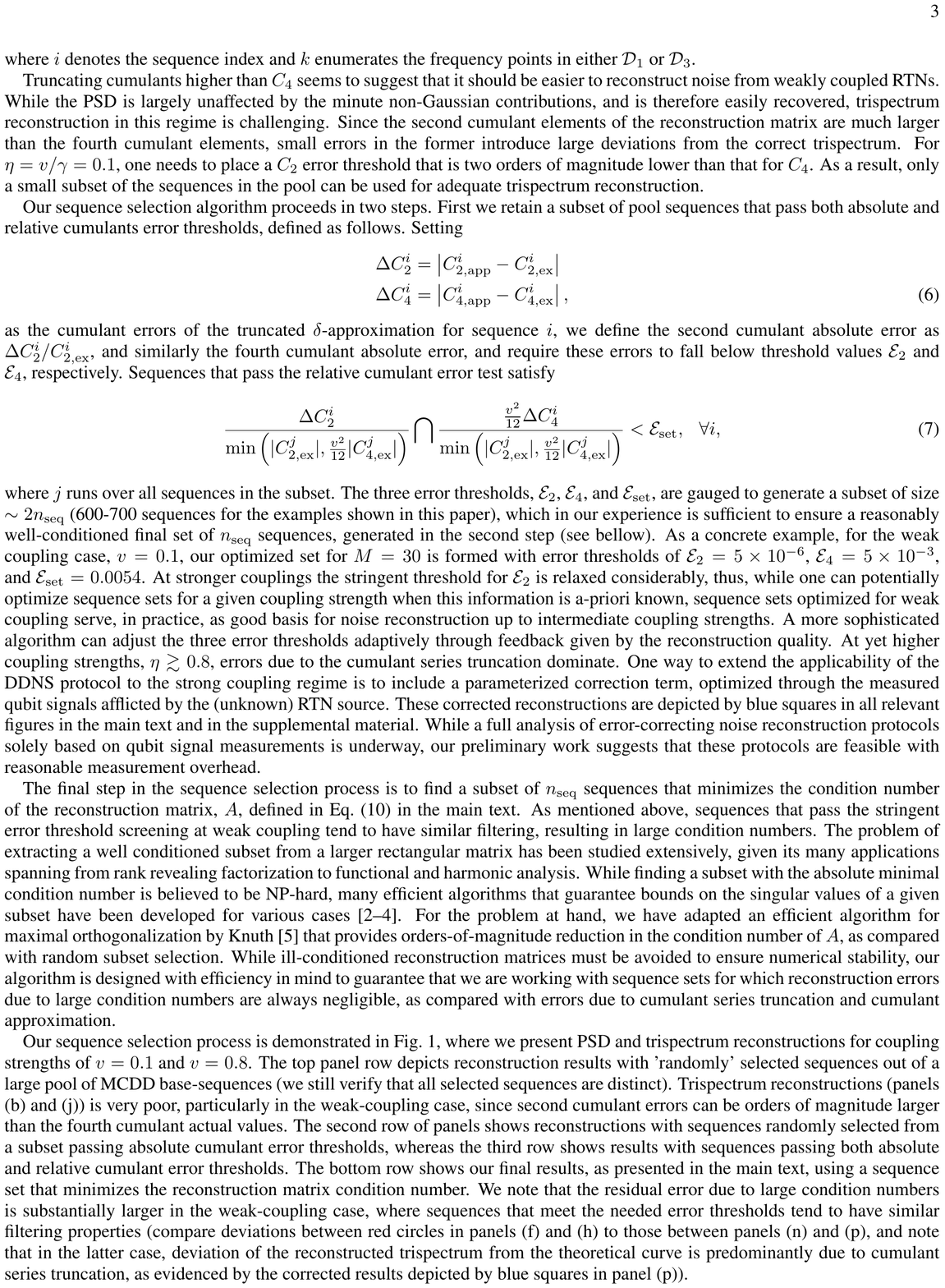}
\hspace*{-2cm}
\includegraphics{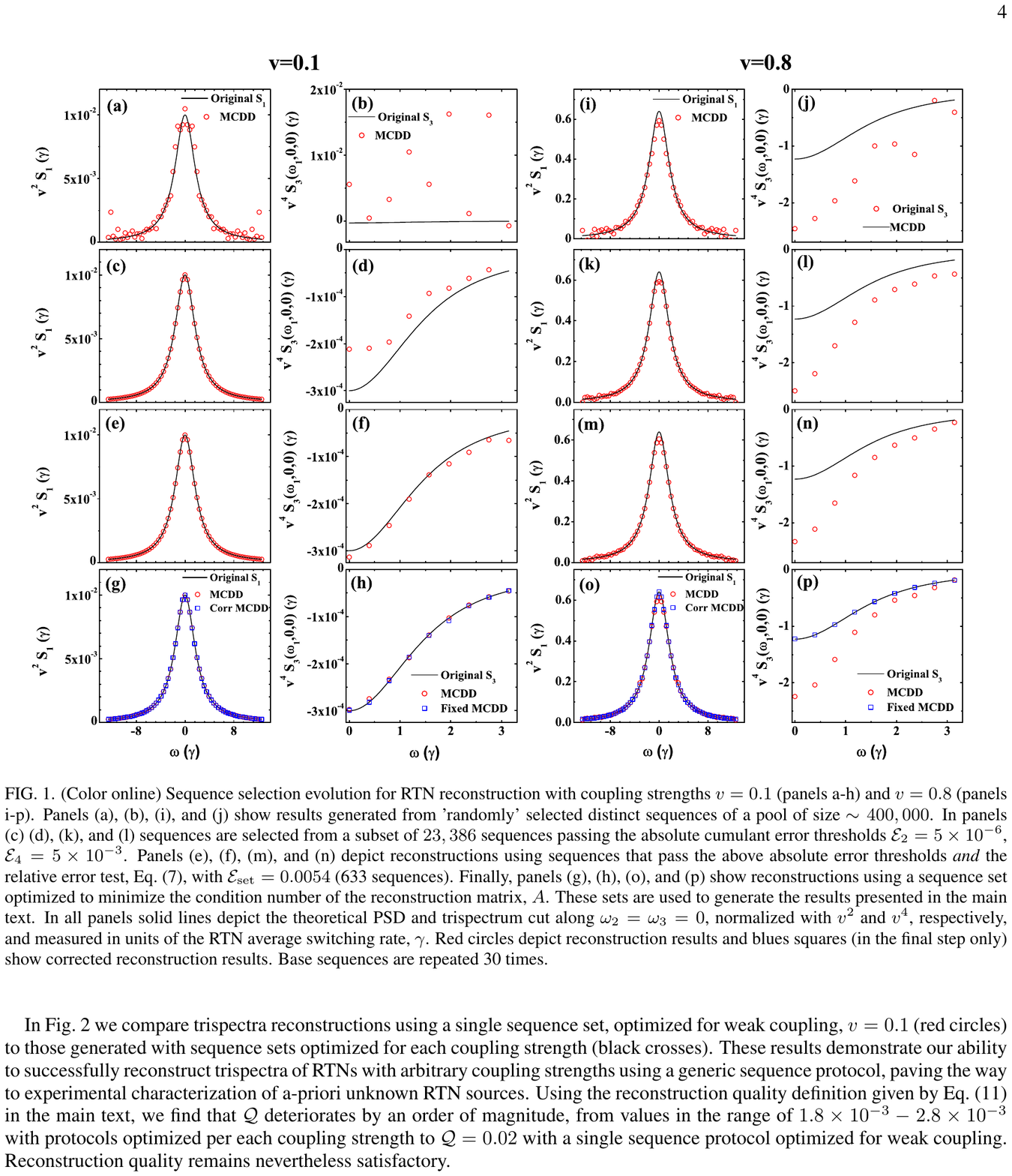}
\hspace*{-2cm}
\includegraphics{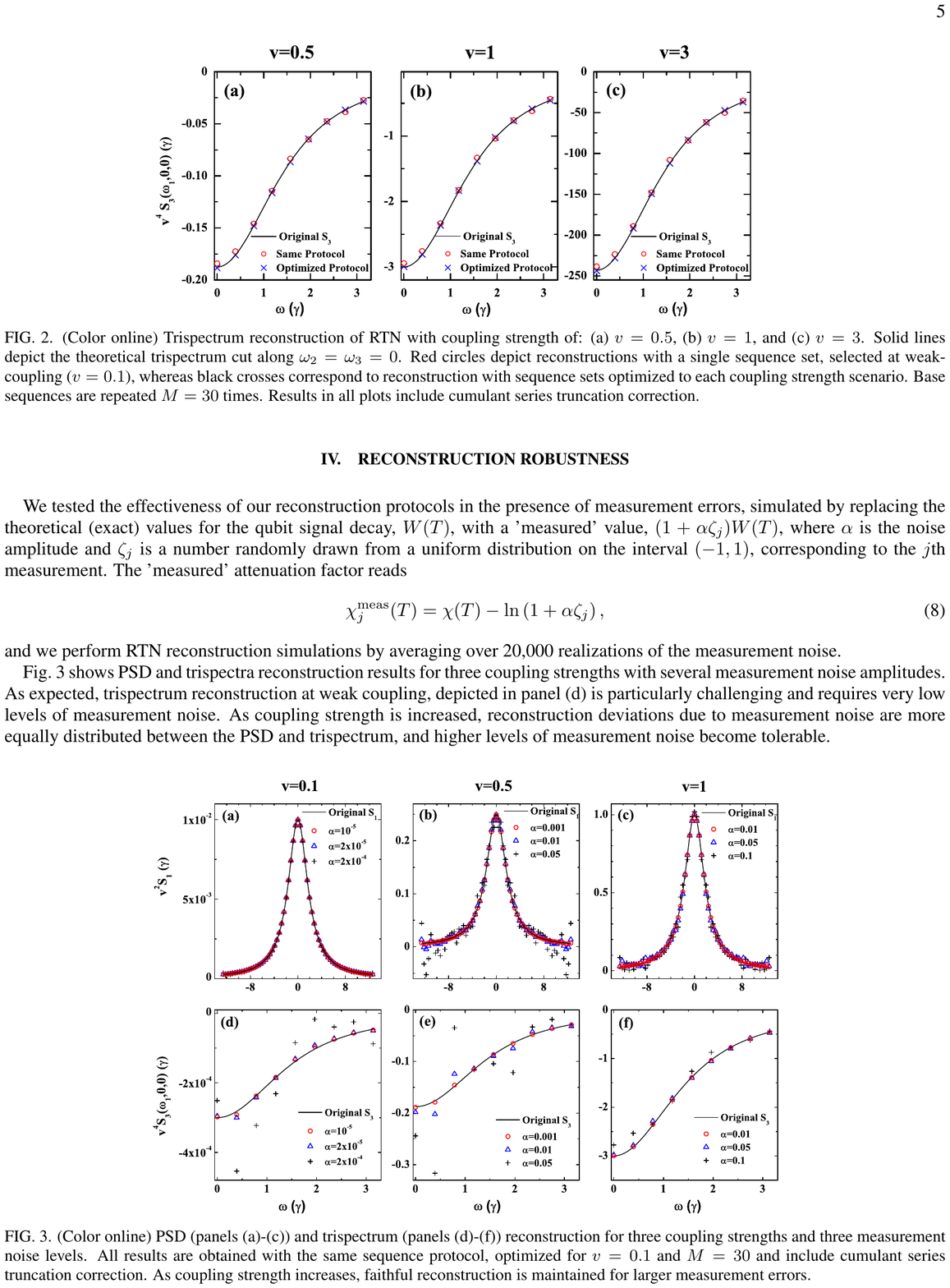}
\hspace*{-2cm}
\includegraphics{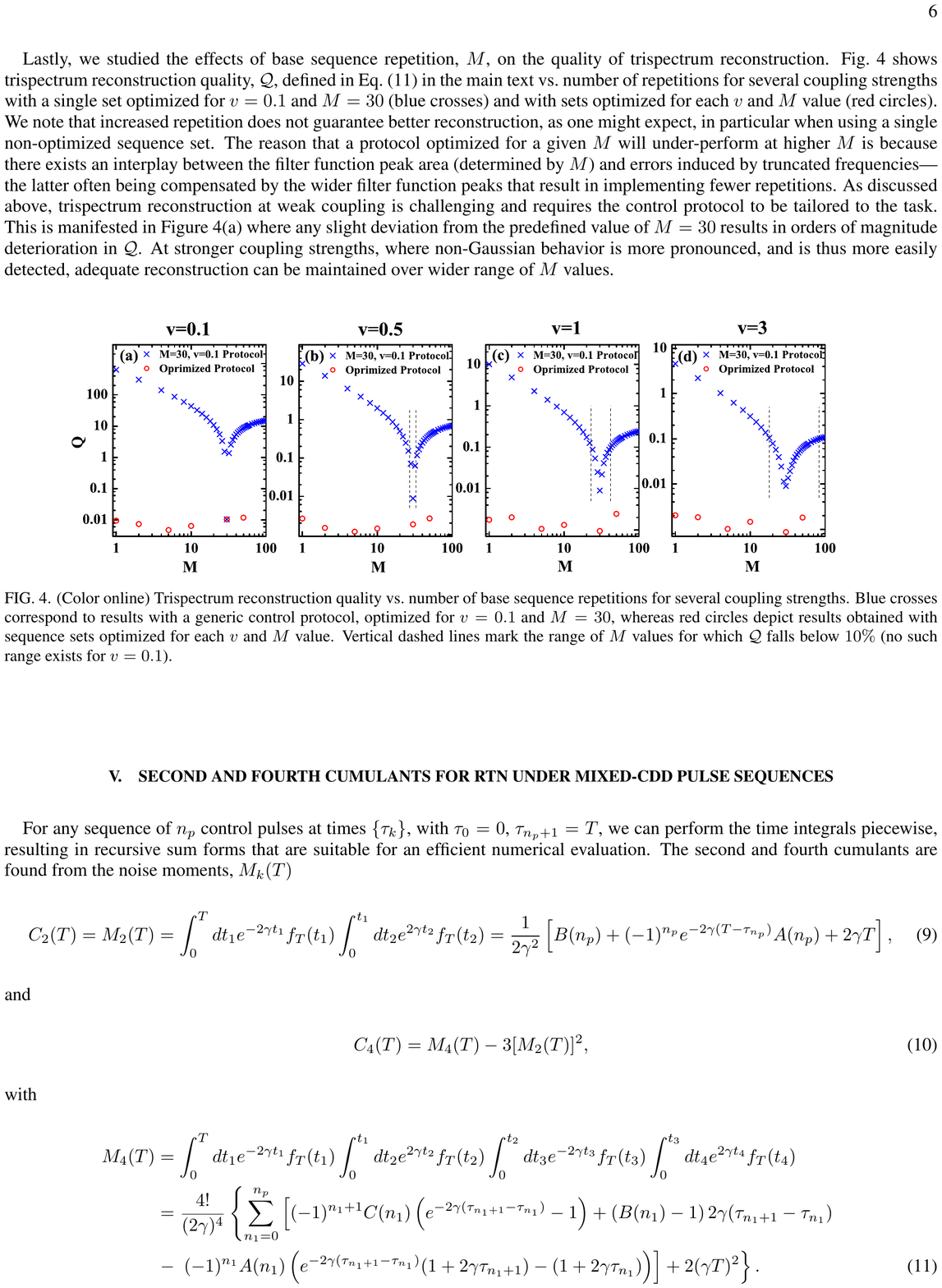}
\hspace*{-2cm}
\includegraphics{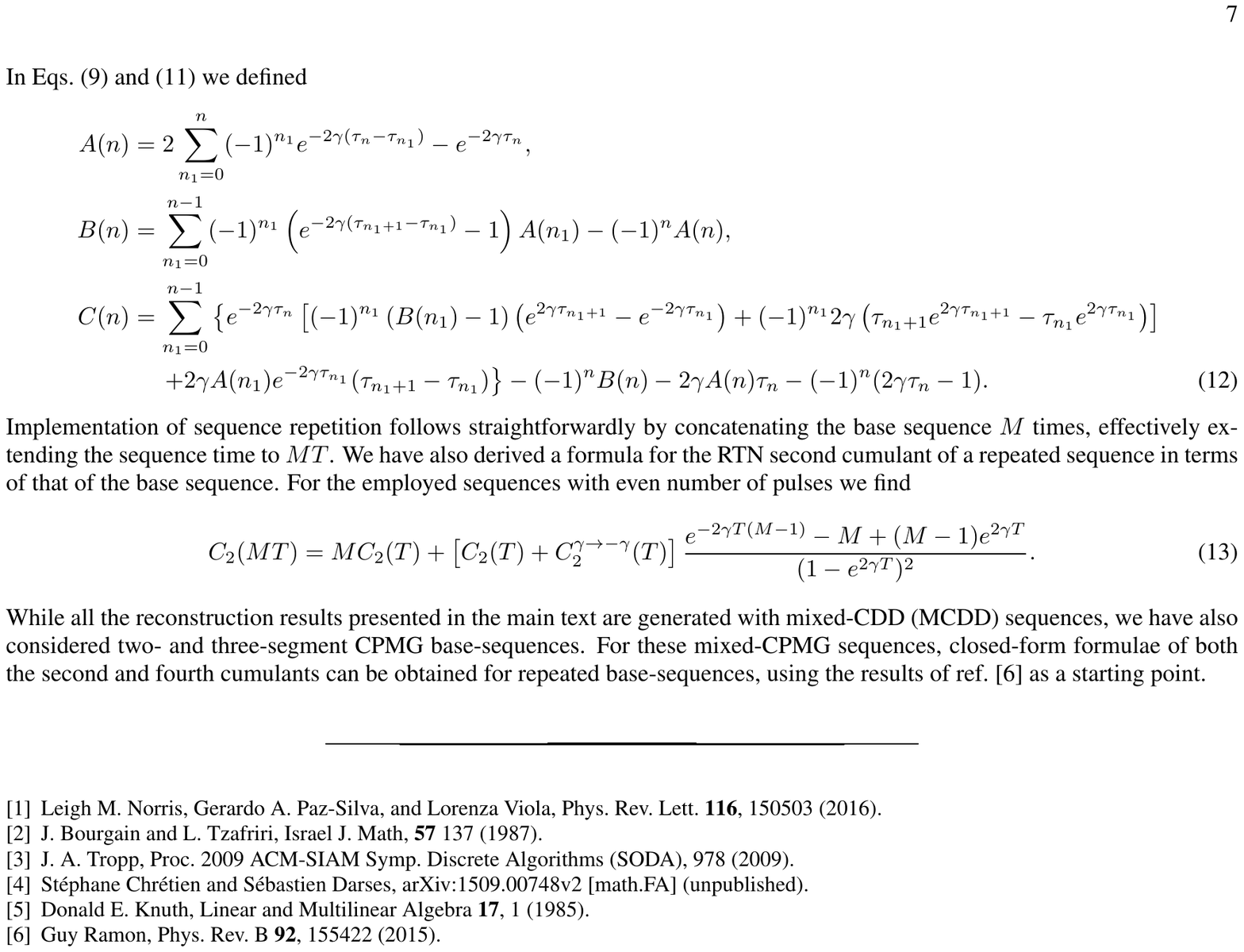}

\end{document}